\documentclass[aoas,MSNbibl,nameyear,dvips]{arximspdf}
\usepackage{algorithmic}
\usepackage{algorithm}
\usepackage{graphicx}

\doi{10.1214/12-AOAS543} 
\volume{6}
\issue{3}
\pubyear{2012}
\firstpage{1280}
\lastpage{1305}

\makeatletter

\setattribute{abstract}{skip}{25\p@}
\setattribute{frontmatter} {skip} {0\p@ plus 3\p@ minus 3\p@}

\newtheorem{theorem}{Theorem}
\newtheorem{lemma}[theorem]{Lemma}

\newproclaim{definition}{Definition}
\newcommand{\eqref}[1]{(\ref{#1})}
\makeatother

\begin{document}
\begin{frontmatter}

\title{Statistical methods for tissue array images---algorithmic
scoring and co-training}
\runtitle{TACOMA}

\begin{aug}
\author[A]{\fnms{Donghui} \snm{Yan}\ead[label=e1]{dyan@fhcrc.org}},
\author[A]{\fnms{Pei} \snm{Wang}\ead[label=e2]{pwang@fhcrc.org}\thanksref{t1}},
\author[B]{\fnms{Michael} \snm{Linden}\ead[label=e3]{linde013@umn.edu}},
\author[C]{\fnms{Beatrice}~\snm{Knudsen}\ead[label=e4]{Beatrice.Knudsen@cshs.org}}
\and
\author[A]{\fnms{Timothy} \snm{Randolph}\corref{}\ead[label=e5]{trandolp@fhcrc.org}\thanksref{t1}}
\thankstext{t1}{Supported in part by the following Grants for the
National Institute of Health:
R01CA126205, R01GM082802, U1CA086368, P01CA053996.}
\runauthor{D. Yan et al.}
\affiliation{Fred Hutchinson Cancer Research Center,
Fred Hutchinson Cancer Research Center,
University of Minnesota Medical School,
Cedars-Sinai Medical Center and
Fred Hutchinson Cancer Research Center}
\address[A]{D. Yan\\
P. Wang\\
T. Randolph\\
Biostatistics and Biomathematics Program\\
Fred Hutchinson Cancer Research Center\\
Seattle, Washington 98109\\
USA\\
\printead{e1}\\
\phantom{E-mail: }\printead*{e2}\\
\phantom{E-mail: }\printead*{e5}}
\address[B]{M. Linden\\
Department of Laboratory Medicine\\
\quad and Pathology\\
University of Minnesota Medical School\\
Minneapolis, Minnesota 55455\\
USA\\
\printead{e3}}
\address[C]{B. Knudsen\\
Department of Pathology and\\
\quad Laboratory Medicine\\
Department of Biomedical Sciences\\
Cedars-Sinai Medical Center\\
Los Angeles, California 90048\\
USA\\
\printead{e4}}
\end{aug}

\received{\smonth{1} \syear{2011}}
\revised{\smonth{1} \syear{2012}}

%
\begin{abstract}
Recent advances in tissue microarray technology have allowed
immunohistochemistry to become a powerful medium-to-high throughput
analysis tool, particularly for the validation of diagnostic and
prognostic biomarkers. However, as study size grows, the manual
evaluation of these assays becomes a prohibitive limitation; it
vastly reduces throughput and greatly increases variability and
expense. We propose an algorithm---Tissue Array Co-Occurrence Matrix
Analysis (TACOMA)---for quantifying cellular phenotypes based on
textural regularity summarized by local inter-pixel relationships.
The algorithm can be easily trained for any staining pattern, is
absent of sensitive tuning parameters and has the ability to report
salient pixels in an image that contribute to its score.
Pathologists' input via informative training patches is an important
aspect of the algorithm that allows the training for any specific
marker or cell type. With co-training, the error rate of TACOMA can
be reduced substantially for a very small training sample (e.g.,
with size $30$). We give theoretical insights into the success of
co-training via thinning of the feature set in a high-dimensional
setting when there is ``sufficient'' redundancy among the features.
TACOMA is flexible, transparent and provides a scoring process that
can be evaluated with clarity and confidence. In a study based on an
estrogen receptor (ER) marker, we show that TACOMA is comparable to,
or outperforms, pathologists' performance in terms of accuracy and
repeatability.
\end{abstract}

%

\end{frontmatter}
\newpage
\section{Introduction}
Tissue microarray (TMA) technology was first described by
\citet{WanFF1987} and substantially improved by \citet{Kononen98}
as a high-throughput technology for the assessment of protein
expression in tissue samples.
%
%
\begin{figure}

\includegraphics{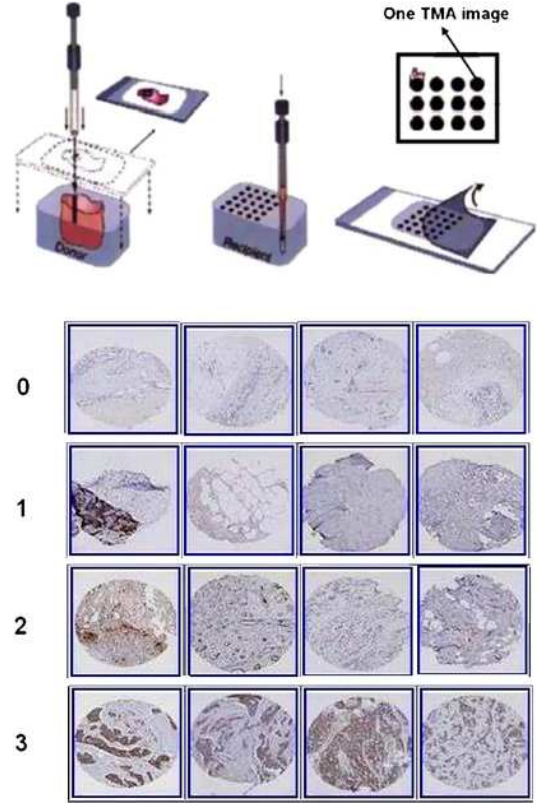}

\caption{Illustration of the TMA
technology and example TMA images. The top
panel is reprinted from Kononen et al. (\protect\citeyear{Kononen98}) by courtesy of the Nature
Publishing Group and it shows the steps involved in how a
TMA image may be produced. The bottom panel displays example TMA
images with the numbers in the left indicating the score of images
in the same row.}\label{figureexpTMAs}
\end{figure}
As shown in the top panel of
Figure~\ref{figureexpTMAs}, the construction of a TMA begins with
cylindrical cores extracted from a donor block of formalin-fixed and
paraffin-embedded tissues. The cores are transferred to the grid of
the recipient block. This grid is generated by punching cylindrical
holes at equal distance into a precast rectangular mold of solid
paraffin wax. Once all the holes are filled with donor cores, the
block is heated to fuse the cores to the wax of the block. Normally,
recipient blocks contain 360 to 480 tissue cores from donor blocks,
often in triplate samples from each block and are thus called tissue
micro arrays (TMA). They are sectioned transversely and each
section is captured on a glass slide, such that slides display
a
cross section of each core in a grid-like fashion. More than 100
slides can be generated from each TMA block for analysis with
a~separate probe. This procedure standardizes the hybridization
process of the probe across hundreds of tissue samples. The use of
TMAs in cancer biology has increased dramatically in recent years
[\citet{CampNeuRimm08}, \citet{Giltnane04}, \citet{Hassan08}, \citet
{Voduc08}] for the rapid
evaluation of DNA, RNA and protein expressions on large numbers of
clinical tissue samples; they remain the most efficient method for
validating proteomics data and tissue biomarkers. We limit our
discussion to high-density immunohistochemistry (IHC) staining, a
method used for the measurement of protein expression, as the most
common method for subcellular localization.

The evaluation of protein expression requires the quantification, or
scoring, of a TMA image. The scores can be used for the validation
of biomarkers, assessment of therapeutic targets, analysis of
clinical outcome, etc. [\citet{Hassan08}]. The bottom panel of
Figure~\ref{figureexpTMAs} gives an example of several TMA images
with scores assigned at a 4-point scale (see
Section~\ref{sectionexperiments} for details).

Although the construction of TMAs has been automated for large-scale
interrogation of markers in tissue samples, several factors limit
the use of the TMA as a high-throughput assay. These include the
variability, subjectivity and time-intensive effort inherent in the
visual scoring of staining patterns
[\citet{CampNeuRimm08}, \citet{Vrolijk03}]. Indeed, a pathologist's score
relies on subjective judgments about colors, textures, intensities,
densities and spatial relationships. As noted in
\citet{Giltnane04}, however, the human eye cannot provide an
objective quantification that can be normalized to a reference. In
general, problems stemming from the subjective and inconsistent
scoring by pathologists are well known and have been highlighted by
several studies
[\citet{Bentzen08}, \citet{Berger05}, \citet{Divito05},
\citet{Kirkegaard06}, \citet{Thomson01}, \citet{Walker06}].
Thus, as study size grows, the value of TMAs in a rigorous
statistical analysis may actually decrease without a consistent and
objective scoring process.

These concerns have motivated the recent development of a variety of
tools for automated scoring, ranging from sophisticated image
enhancement tools, tissue segmentation to computer-assisted
pathologist-based scoring. Many are focused on a particular
cellular pattern, with HER2 (exhibiting nuclear staining) being the
most commonly targeted marker; see, for example,
\citet{Hall08}, \citet{Joshi07}, \citet{Masmoudi09}, \citet{Skaland08},
\citet{Tawfik05}. For a
survey of commercial systems, we refer to \citet{Mulrane08} or
\citet{Rojo09}, and also the review by \citet{Cregger06} which
acknowledges that, given the rapid changes in this field, this
information may become outdated as devices are abandoned, improved
or newly developed. A~property of\vadjust{\goodbreak} most automated TMA scoring
algorithms is that they rely on various forms of background
subtraction, feature segmentation and thresholds for pixel
intensity. Tuning of these algorithms can be difficult and may
result in models sensitive to several variables, including staining
quality, background antibody binding, counterstain intensity, and
the color and hue of chromogenic reaction products used to detect
antibody binding. Moreover, such algorithms typically require tuning
from the vendors with parameters specific to the markers' staining
pattern (e.g., nuclear versus cytoplasmic), or even require a
dedicated person for such a system.

To address the further need for scoring of TMAs in large biomarker
studies, we propose a framework---called Tissue Array Co-Occurrence
Matrix Analysis (TACOMA)---that is trainable to any staining pattern
or tissue type. By seeking texture-based patterns invariant in the
images, TACOMA does not rely on intensity thresholds, color filters,
image segmentation or shape recognition. It recognizes specific
staining patterns based on expert input via a preliminary set of
image patches. In addition to providing a~score or categorization,
TACOMA allows to see which pixels in an image contribute to its
score. This clearly enhances interpretability and confidence in the
results.

It should be noted that TACOMA is not designed for clinical
diagnosis but rather a tool for use in large clinical studies that
involve a range of potential biomarkers. Since many thousands of
samples may be required, the cost and time required for
pathologist-based scoring may be prohibitive and so an efficient
automated alternative to human scoring can be essential. TACOMA is a
framework for such a purpose.

An important concern in biomedical studies is that of the limited
training sample size.\setcounter{footnote}{1}\footnote{The additional
issue of label noise
was studied elsewhere [\citet{YanGCZ2011}].} The size of the training
set may necessarily be small due to the cost, time or human efforts
required to obtain them. We adopt co-training [\citet{Yarowsky1995},
\citet{BlumMitchell1998}] in the context of TACOMA to substantially reduce
the training sample size. We explore the thinning of the feature set
for co-training when a ``natural'' split is not readily available but
the features are fairly redundant, and this is supported by our
theory that a thinned slice carries about the same classification
power as the whole feature set under some conditions.

The organization of the remainder of this paper is as follows. We
describe the TACOMA algorithm in Section~\ref{sectiontacoma}, this
is followed by a discussion on co-training to reduce the training
sample size with some theoretical insights on the thinning scheme in
Section~\ref{sectioncoTraining}. Then in
Section~\ref{sectionexperiments}, we present our experimental
results. We conclude with a discussion in
Section~\ref{sectionconclusion}.

\section{The TACOMA algorithm}
\label{sectiontacoma} 
\begin{figure}
\centering
\begin{tabular}{@{}c@{}}

\includegraphics[scale=0.97]{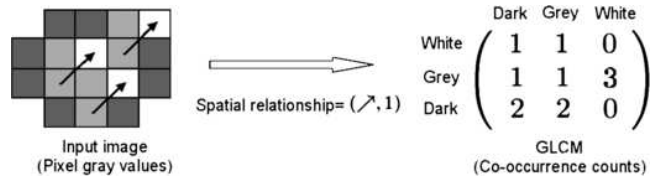}
 \\
\footnotesize{(a)}\\

\includegraphics[scale=0.97]{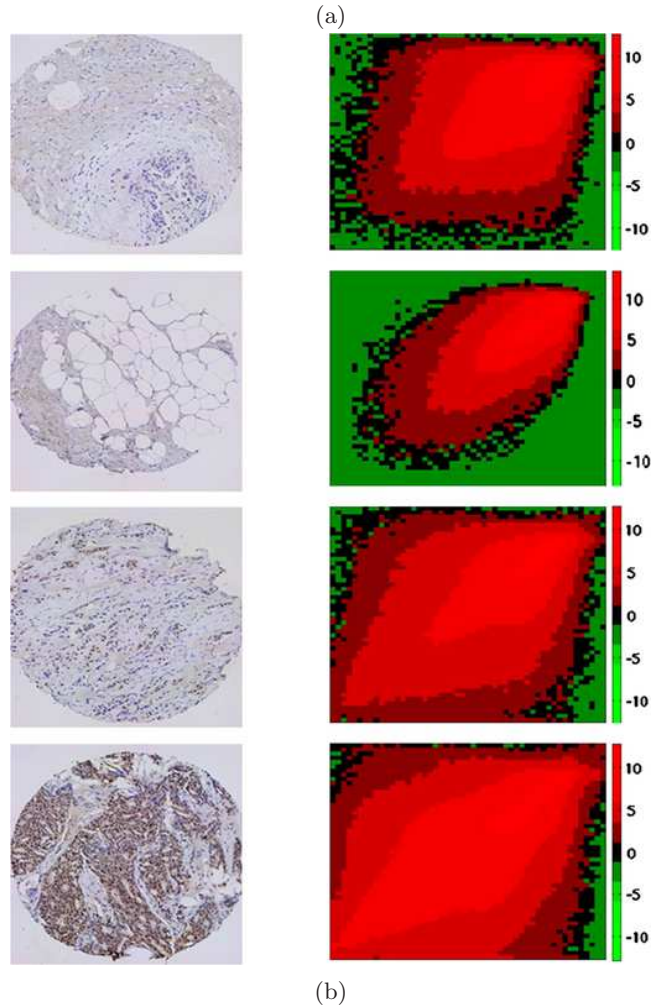}
\\
\footnotesize{(b)}\\
\end{tabular}
\caption{Example images and their
GLCMs. \textup{(a)} Generating the GLCM from an image. This toy
``image'' (left) has $3$ gray levels, $\{\mathit{Dark}, \mathit{Grey}, \mathit{White}\}$. Here,
under the spatial relationship $(\nearrow,1)$, the transition from
$\mathit{Grey}$ to $\mathit{White}$ (indicated by $\nearrow$) occurs three times;
accordingly, the entry of the GLCM corresponding to the Grey row and
White column has a value of $3$. \textup{(b)} Example TMA images.
Images of a tissue sample (left panel) and the Heatmap (right
panel) of their GLCM (in log scale). The GLCM matrices are all $51
\times51$; each GLCM cell in the heatmap indicates the frequency of
the corresponding transition. The color scale is illustrated by the
color bar on the right.} \label{figureglcms}
\end{figure}

The primary challenge TACOMA addresses is the
lack of easily-quantified criteria for scoring: features of interest
are not localized in position or size.\vadjust{\goodbreak} Moreover, within any region
of relevance---one containing primarily cancer cells---there is no
well-defined (quantifiable) shape that characterizes a pattern of
staining (see, e.g., the bottom panel of
Figure~\ref{figureexpTMAs} for an illustration). The key insight
that underlies TACOMA is that, despite the heterogeneity of TMA
images, they exhibit strong statistical regularity in the form of
visually observable textures or staining pattern [see, e.g.,
Figure~\ref{figureglcms}(b)]. And, with the guidance of
pathologists, TACOMA can be trained for this pattern regardless of
the cancer cell type (breast, prostate, etc.) or marker type (e.g.,
nucleus, cytoplasmic, etc.).

TACOMA captures the texture patterns exhibited by TMA images through
a matrix of counting statistics, the Gray Level Co-occurrence Matrix
(GLCM). Through a small number of representative image patches,
TACOMA constructs a feature mask so that the algorithm will focus on
those biologically relevant features (i.e., a subset of GLCM
entries). Besides scoring, TACOMA also reports salient image pixels
(i.e., those contribute to the scoring of an image) which will be
useful for the purpose of training, comparison of multiple TMA
images, estimation of staining intensity, etc. For the rest of this
section, we will briefly discuss these individual building blocks of
TACOMA followed by an algorithmic description of TACOMA.

%

\subsection{The gray level co-occurrence matrix}
The GLCM was originally proposed by \citet{haralick1979} and has
proven successful in a variety of remote-sensing applications
[\citet{ASPRAS2006}]. The GLCM, of an image, is a matrix whose entries
count the frequency of transitions between pixel intensities across
neighboring pixels with a particular spatial relationship; see
Figure~\ref{figureglcms}. The description here is essentially
adopted from \citet{ASPRAS2006}. We start by defining the spatial
relationship between a~pair of pixels in an image.

\begin{definition*} A spatial relationship has two elements, the
direction and the distance of interaction. The set of all possible
spatial relationships is defined as
\begin{eqnarray*}
\Re&=&\mathit{D} \otimes\mathit{L} \\
&=& \{\nearrow, \searrow, \nwarrow, \swarrow, \downarrow,
\uparrow, \rightarrow, \leftarrow\} \otimes\{1,\ldots,d\},
\end{eqnarray*}
where $D$ is the set of potential directions and $L$ is the distance
of interaction between the pair of pixels involved in a spatial
relationship. The distance of interaction is the minimal number of
steps required to move from one pixel to the other along a given
direction. The particular spatial relationships used in our
application are $(\nearrow,3)$, $(\searrow,1)$ and $(\nearrow,1)$.
Details about the choice of these spatial relationships can be seen
in Section~\ref{sectionexperiments}.
\end{definition*}

Although the definition of spatial relationships can be extended to
involve more pixels [\citet{ASPRAS2006}], we have focused on pairwise
relationships which appear to be sufficient. Next we define the
GLCM.

\begin{definition*} Let $N_g$ be the number of gray levels in an
image.\footnote{Often before the computing of GLCM, the gray level
of each pixel in an image is scaled linearly from $[1,N_o]$ to
$[1,N_g]$ for $N_o$ the predefined number, typically $256$, of gray
levels in the image.} For a~given image (or a patch) and a fixed
spatial relationship $\sim\,\in\Re$, the GLCM is defined as
\[
\begin{tabular}{p{340pt}@{}}
A $N_g \times N_g$ matrix such that its $(a,b)$-entry
counts the number of pairs of pixels, with gray values $a
\mbox{ and } b$, respectively, having a spatial relationship $\sim$,
for $a, b \in\{1,2,\ldots,N_g\}$.
\end{tabular}
\]

This definition is illustrated in Figure~\ref{figureglcms}(a) with
more realistic examples in Figure~\ref{figureglcms}(b).
Figure~\ref{figureglcms}(b) gives a clear indication as to how the
GLCM distinguishes between TMA images having different staining
patterns.
\end{definition*}

Our use of the GLCM is nonstandard in that we do not use any of the
common scalar-valued summaries of a GLCM [see \citet{haralick1979}
and \citet{Conners80}], but instead employ the entire matrix (with
masking) in a classification algorithm [see also
\citet{ASPRAS2006}]. A~GLCM may have a large number of entries,
typically thousands, however, the exceptional capability of Random
Forests [\citet{RF}] in feature selection allows us to directly use
all (or a masked subset of) GLCM entries to determine a final score
or classification.

\subsection{Image patches for domain knowledge}
\begin{figure}

\includegraphics{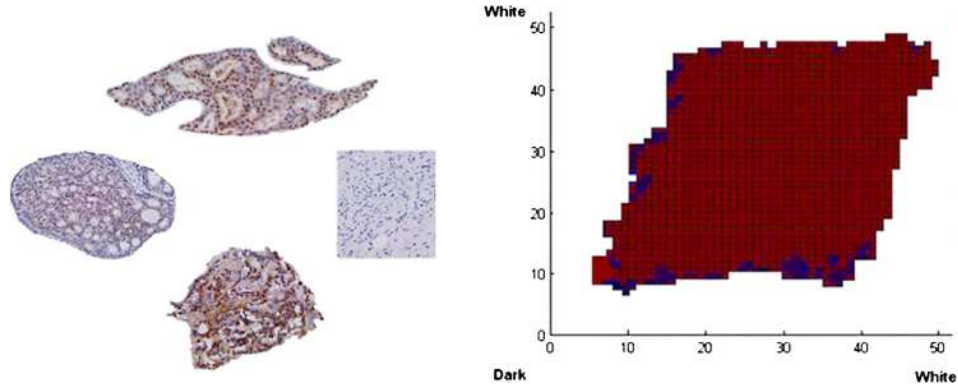}

\caption{Representative image patches
and the induced feature mask. Four pathologist-chosen patches
(left panel) and the feature mask as determined by all patches
(right panel, see algorithmic description of TACOMA). Nonwhite
entries in this matrix indicate the corresponding GLCM entries to be
used in scoring. Note that one and only one feature mask is required
throughout.} \label{figureimgPatch}
\end{figure}

In order to incorporate prior knowledge about the staining pattern,
we mask the GLCM matrix so that the scoring will focus on
biologically pertinent features. The masking is realized by first
choosing a set of image patches representing regions that consist
predominantly of cancer cells and are chosen to represent the
staining patterns; see Figure~\ref{figureimgPatch}. The collection
of GLCMs from these patches are then used to define a template of
``significant entries'' (cf. TACOMA algorithm in
Section~\ref{sectionalgorithmTACOMA}) for all future GLCMs: when
the GLCM of a new image is formed, only the entries that correspond
to this template are retained. This masking step enforces the idea
that features used in a classifier should not be based on stromal,
arterial or other nonpertinent tissue which may exhibit
nonspecific or background staining. Note that only one small set of
image patches is required; these image patches are used to produce
a common feature mask which is applied to all images in both
training and scoring.

In this fashion, feature selection is initiated by expert biological
knowledge. This manner of feature selection involves little human
effort but leads to substantial gain in both interpretability and
accuracy. The underlying philosophy is that no machine learning
algorithms surpass domain knowledge. Since by using image patches we
do not indicate which features to select but instead specify their
effect, we achieve the benefits of a manual-based feature selection
but avoid its difficulty. This is a novel form of nonparametric, or
implicit, feature selection which is applicable to settings beyond
TMAs.

%

%
\begin{table}[b]
\caption{Performance comparison of RF, SVM and Boosting of a
naive Bayes classifier. The~result for RF is adopted from
Section~\protect\ref{subsectionexpScore}. For SVM, we vary the choice
of~the~kernel from \{Gaussian, polynomial, sigmoid\} with the best
tuning parameters for~$N_g \in\{7, 9, 13, 26, 37, 51, 64, 85\}$ and
the best result is reported. For boosting, the~best~result is
reported by varying $N_g \in\{7, 9, 13, 26, 37, 51, 64, 85\}$ and
the~number~of~boosting iterations from $\{1, 2,3,5,10,50,100\}$}
\label{tablecomparisonRFSVMBoost}
\begin{tabular*}{120pt}{@{\extracolsep{\fill}}lc@{}}
\hline \textbf{Classifier} &\textbf{Accuracy} \\
\hline
RF &$78.57\%$ \\
SVM &$65.24\%$ \\
Boosting &$61.28\%$ \\
\hline
\end{tabular*}
\end{table}

\subsection{Random forests}\label{subsectionsalientPixels} 
\begin{figure}

\includegraphics{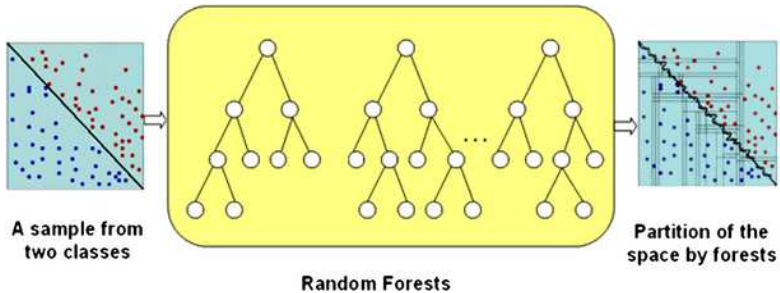}

\caption{Random Forests classification. In this
illustration the data points reside in a~unit square (left panel).
The two classes are indicated by red and blue dots. The true
decision boundary is the diagonal line shown. RF (center panel)
grows many trees. Each tree corresponds to a recursive partition of
the data space. These partitions are represented in the right panel by
a sequence of horizontal and vertical lines; the data space shown
here is partitioned by many instances. The RF classifier eventually
leads to a decision boundary (solid black curve) for this two-class
classification problem.} \label{figurerf}
\end{figure}

TACOMA uses
Random Forests (RF)
[\citet{RF}] as the underlying classifier. RF was proposed by Breiman
and is considered one of the best classifiers in a high-dimensional
setting [\citet{caruanaKY2008}]. In our experience, RF achieves
significantly better performance than SVM and Boosting on the TMA
images we use (see Table~\ref{tablecomparisonRFSVMBoost}).
Additionally \citet{HolmesKapelner2009} argue that RF is superior to
others in dealing with tissue images. The fundamental building block
of RF is a tree-based classifier which can be nonstable and
sensitive to noise. RF takes advantage of such instability and
creates a~strong ensemble by bagging a large number of trees
[\citet{RF}]. Each individual tree is grown on a bootstrap sample from
the training set. For the splitting of tree nodes, RF randomly
selects a number of candidate features or linear combinations of
features and splits the tree node with the one that achieves the
most reduction in the node impurity as defined by the Gini index [or
other measures such as the out of bag (oob) estimates of
generalization error] defined as follows:
%
\begin{equation}
\label{eqgini} \phi(\mathbf{p}) = \sum_{i=1}^C p_i(1-p_i),
\end{equation}
where $\mathbf{p}=(p_1,\ldots,p_C)$ denotes the proportion of examples from
different classes and $C$ is the number of different classes. RF
grows each tree to the maximum and no pruning is required. For an
illustration of RF, see Figure~\ref{figurerf}.

To test a future example $X$, let $X$ fall from each tree for which
$X$ receives a vote for the class of the terminal node it reaches.
The final class membership of~$X$ is obtained by a majority vote
over the number of votes it receives for each class. The features
are ranked by their respective reduction of node impurity as
measured by the Gini index. Alternatives include the
permutation-based measure, that is, permute variables one at a time
and then rank according to the respective amount of decrease in
accuracy (as estimated on oob observations).

%

\subsection{Salient pixels detection}
A valuable property of TACOMA is its ability to report salient
pixels in an image that determine its score (see
Figure~\ref{figuresalientfeaturepixels}). This property is based
on\vadjust{\goodbreak}
a correspondence between the position of pixels in an image and
entries in its GLCM, and made possible by the remarkable
variable-ranking capability of RF. Here we use the importance
measure (Gini index-based) provided by RF to rank the variables
(i.e., entries of the GLCM) and then collect relevant image pixels
associated with the important entries. 

Since each entry of a GLCM is a counting statistic involving pairs
of pixels, we can associate the $(a,b)$-entry of a GLCM with those
pixels that make up this GLCM entry. The set of image pixels that
are associated with the $(a,b)$-entry of a GLCM is formally
represented as
\[
\mathcal{G}_{a,b}=\{x,y\dvtx x \sim y, I(x)=a,I(y)=b\}.
\]
In the above representation, $x$ and $y$ represent the position of
image pixels and we treat an image $I$ as a map from the position of
an image pixel to its gray value. Note that not all pairs of pixels
with $x \sim y$ such that $I(x)=a,I(y)=b$ correspond to salient
spots in a TMA image. However, if the $(a,b)$-feature is ``important''
(e.g., as determined by RF), then typically most pixels in the set
$\mathcal{G}_{a,b}$ are relevant.

Whereas RF appears to be a black box---taking a large number of GLCM
features and producing a score---salient pixels provide a quick peek
into its internals. Effectively, RF works in roughly the same manner
as a pathologist, that is, they both use salient pixels to score the
TMA images; the seemingly mysterious image features are merely a
form of representation for use by a~computer algorithm.

\subsection{An algorithmic description of TACOMA}
\label{sectionalgorithmTACOMA} Denote the training sample by
$(I_1,Y_1),\ldots,(I_n,Y_n)$ where $I_i$'s are images and $Y_i$'s are
scores. Additionally, assume there is a set of $L$ ``representative''
image patches. The training of TACOMA is described as
Algorithm~\ref{algorithmtacoma}.

\begin{algorithm}[t]
\caption{The training in TACOMA}\label{algorithmtacoma}
\begin{algorithmic}[1]
\FOR {$i=1$ \textbf{to} $L$} 
\STATE compute the GLCM for the $i${th} image patch and denote by
$Z_i$;
\STATE $M_i \gets$ the index set of $Z_i$ that survive thresholding
at level $\tau_i$; 
\ENDFOR 
\STATE $M \gets \bigcup_{j=1}^L M_j$; 
\FOR {$k=1$ \textbf{to} $n$} 
\STATE compute the GLCM of image $I_k$ and keep only entries in
index set $M$; 
\STATE denote the resulting matrix by $X_k$; \ENDFOR 
\STATE Feed $\bigcup_{l=1}^n \{(X_l,Y_l)\}$ into the RF
    classifier and obtain a classification rule.
\end{algorithmic}
\end{algorithm}

\begin{figure}[b]
\vspace*{3pt}
\includegraphics{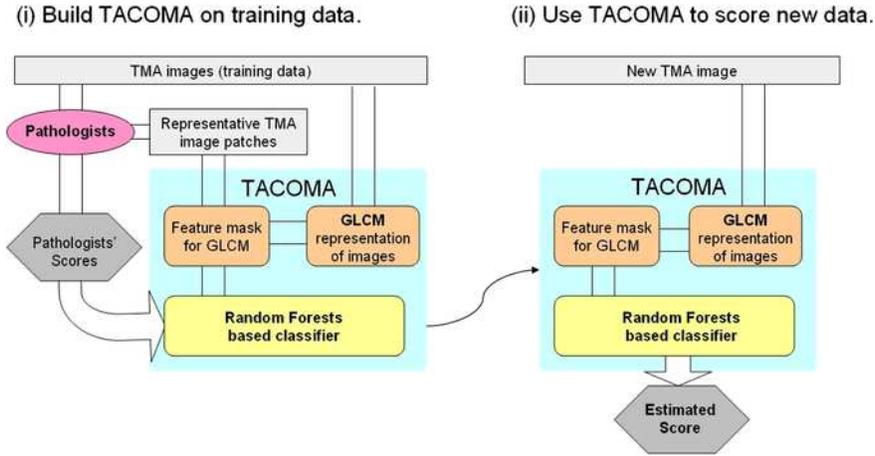}

\caption{TACOMA illustrated. The
left and right panels illustrate, respectively, model training and
the use of the model on future data.} \label{figurefChart}
\end{figure}

In the above description, $\tau_i$ is chosen as the median of
entries of matrix~$Z_i$ for $i=1,\ldots,L$. Then, for a new image,
TACOMA will: (i) derive the GLCM matrix; (ii) select the entries
with indices in $M$; (iii) apply the trained classifier on the
selected entries and output the score. The training and scoring with
TACOMA are illustrated in Figure~\ref{figurefChart}.

\section{Co-training with RF}
\label{sectioncoTraining} 
\begin{figure}

\includegraphics{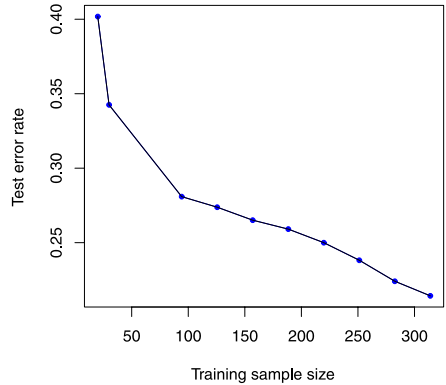}

\caption{Error rate of TACOMA as the training sample size
varies. There are $328$ TMA images in the test sample.}\label{figureerrSample}\vspace*{3pt}
\end{figure}

The sample size is an important issue in
the scoring of TMA images, mainly because of the high cost and human
efforts involved in obtaining a large sample of high quality labels.
For instance, it may take several hours for a well-trained
pathologist to score $100$ TMA images. Unfortunately, it is often
the case that the classification performance drops sharply when the
training sample size is reduced. For example,
Figure~\ref{figureerrSample} shows the error rate of TACOMA when
the sample size varies. Our aim is to achieve reasonable accuracy
for small sample size and co-training is adopted for this purpose.

%

%

%
\begin{figure}[b]
\vspace*{3pt}
\includegraphics{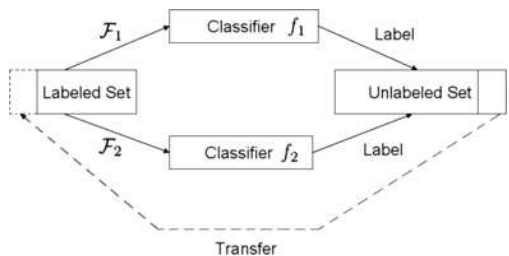}

\caption{An illustration of co-training.}
\label{figurecoTrain}
\end{figure}

Co-training was proposed in the landmark papers by
\citet{Yarowsky1995} and \citet{BlumMitchell1998}. It is an
effective way in training a model with an extremely small labeled
sample and has been successfully applied in many applications
[\citet{BlumMitchell1998}, \citet{NigamGhani2000}]. The idea is to
train two
separate classifiers (called coupling classifiers) each on a
different set of features using a small number of labeled examples.
Then the two classifiers iteratively transfer those confidently
classified examples, along with the assigned label, to the labeled
set. This process is repeated until all unlabeled examples have been
labeled. For an illustration of the idea of co-training, see
Figure~\ref{figurecoTrain}. Co-training is relevant here due to the
natural redundancy that exists among features that are based on
GLCMs corresponding to different spatial relationships.\vadjust{\goodbreak}

A learning mode that is closely related to co-training is
self-learning [\citet{NigamGhani2000}], where a single classifier is
used in the ``$\mathit{label} \rightarrow \mathit{transfer} \rightarrow \mathit{label}$'' loop
(cf. Figure~\ref{figurecoTrain}). However, empirical studies have
shown that co-training is often superior [\citet{NigamGhani2000}]; the
intuition is that co-training allows the two coupling classifiers
to progressively expand the ``knowledge boundary'' of each other which
is absent in self-learning.

Previous works in co-training use almost exclusively Expectation
Maximization or Naive Bayes based classifiers where the posterior
probability serves as the ``confidence'' required by co-training.
Here we use RF [\citet{RF}] where the~margin (to be defined shortly)
provided by RF is used as a ``natural'' proxy for the ``confidence.''
The margin is defined through the votes received by an observation.
For an observation $x$ in the test set, let the number of votes it
receives for the $i${th} class be denoted by $N_i(x),i=1,\ldots,C$,
where $C$ is the number of classes. The \textit{margin} of $x$ is
defined as
\[
\max_{i \in\{1,\ldots,C\}}N_i(x)- \mathop{\mathrm{second}}_{j \in\{1,\ldots,C\}}
 N_j(x),\vadjust{\goodbreak}
\]
where $\mathit{second}$ in the above indicates the second-largest element in
a list.

To give an algorithmic description of co-training, let the two
subsets of features be denoted by $\mathcal{F}_1$ and
$\mathcal{F}_2$, respectively. Let the set of labeled and unlabeled
examples be denoted by $\mathcal{L}$ and $\mathcal{U}$,
respectively. The co-training process proceeds as
Algorithm~\ref{algorithmco-training}. The final error rate and
class membership are determined by a fixed coupling classifier, say,
$f_1$. We set $m_1=m_2=2$ in our experiments according to
\citet{BlumMitchell1998}.

\begin{algorithm}[t]
\caption{The co-training algorithm} \label{algorithmco-training}
\begin{algorithmic}[1]
\WHILE{the set $\mathcal{U}$ is not empty}
\FOR{$k=1,2$} 
\STATE Train RF classifier $f_k$ on labeled examples from
$\mathcal{L}$ using feature set $\mathcal{F}_k$; 
\STATE Classify examples in the set $\mathcal{U}$ with $f_k$; 
\STATE Under $f_k$, calculate the margin for each observation in
$\mathcal{U}$;
\STATE pick $m_k$ observations, $x_1^{(k)}, \ldots, x_{m_k}^{(k)}$,
with the largest margins;
\ENDFOR
\STATE$\mathcal{L} \gets\mathcal{L} \cup
\{x_1^{(1)},\ldots,x_{m_1}^{(1)},x_1^{(2)},\ldots,x_{m_2}^{(2)}\}$;
\STATE$\mathcal{U} \gets\mathcal{U} \setminus
\{x_1^{(1)},\ldots,x_{m_1}^{(1)},x_1^{(2)},\ldots,x_{m_2}^{(2)}\};$
\ENDWHILE
\end{algorithmic}
\end{algorithm}
%

\subsection{Feature split for co-training}
\label{sectionfeatureSplit} Co-training requires two subsets of
features (or a feature split). However, co-training algorithms
rarely provide a~recipe for obtaining these feature splits. There
are several possibilities one can explore.

The first is called a ``natural'' split, often resulting from an
understanding of the problem structure. A rule of thumb as to what
constitutes a~natural split is that each of the two feature subsets
alone allows one to construct an acceptable classifier and that the
two subsets somehow complement each other (e.g., conditional
independence given the labels). Fortunately, TMA images represented
in GLCM's naturally have such properties. For a~given problem, often
there exist several \textit{spatial relationships} [e.g.,
$(\nearrow,3)$ and \mbox{$(\searrow,1)$} for TMA images studied in this
work], with each inducing a~GLCM sufficient to construct a classifier
while the ``dependence'' among the induced GLCM's is usually low.
Thus, it is ideal to apply co-training on TMA images using such
natural splits.

When there is no natural split readily available, one has to find
two proper subsets of features. One way is via random splitting.
Co-training via a random split of features was initially considered
by \citet{NigamGhani2000} but has since been largely overlooked in
the machine learning literature. Here we extend the idea of random
splits to ``thinning,'' which is more flexible and potentially may
lead to a better co-training performance. Specifically, rather than
randomly splitting the original feature set
$\mathcal{F}=\{1,\ldots,p\}$ into two halves, we select two disjoint
subsets of $\mathcal{F}$ with size not necessarily equal but
nonvanishing compared to $p$. This way of feature splitting leads
to feature subsets smaller than $\mathcal{F}$, hence the name
``thinning.'' One concrete implementation of this is to divide
$\mathcal{F}$ into a number of, say, $J$, equal-sized partitions
(each partition is also called a thinned slice of $\mathcal{F}$). In
the following discussion, unless otherwise stated, thinning always
refers to this concrete implementation. It is clear that this
includes random splits as a special case. Thinning allows one to
construct a self-learning classifier (the features are taken from one
of the $J$ partitions), co-training (randomly pick $2$ out of $J$
partitions) and so on. For a given problem, one can explore various
alternatives associated with thinning but here we shall focus on
co-training.

The extension of random split to thinning may lead to improved
co-training performance, as thinning may make features from
different partitions less dependent and meanwhile well preserves the
classification power in a high-dimensional setting when there is
sufficient redundancy among features (see
Section~\ref{sectioncotrainingTheory}). The optimal number of
partitions can be selected by heuristics such as the kernel
independence test [\citet{BachJordan2003}, \citet{GrettonFTSSS2006}],
which we
leave for future work.

\subsection{Some theoretical insights on thinning}
\label{sectioncotrainingTheory} According to
Blum and Mit\-chell (\citeyear{BlumMitchell1998}), one essential ingredient of co-training is
the ``high'' confidence of the two coupling classifiers in labeling
the unlabeled examples. This is closely related to the strength of
the two coupling classifiers which is in turn determined by the
feature subsets involved. In this section, we study how much a
thinned slice of the feature set $\mathcal{F}$ preserves its
classification power. Our result provides insight into the nature of
thinning and is interesting at its own right due to its close
connection to several lines of interesting work [\citet{TinHo1998},
\citet{DasguptaGupta2002}] in machine learning (see
Section~\ref{sectionthinningRelated}). We present our theoretical
analysis in Section~\ref{sectionthinningTheorem} and list related
work in Section~\ref{sectionthinningRelated}.
In Supplement A [\citet{refsupp}],
we provide additional simulation results related to our theoretical
analysis.

\subsubsection{Thinning ``preserves'' the ratio of separation}
\label{sectionthinningTheorem} Our theoretical model is the
Gaussian mixture specified as
%
\begin{equation}
\label{eqgm} \Pi\mathcal{N}(\bolds{\mu}_1,\Sigma) +(1-\Pi)
\mathcal{N}(\bolds{\mu}_2,\Sigma),
\end{equation}
where $\Pi\in\{0,1\}$ indicates the label of an observation such
that $\mathbb{P}(\Pi=1)=\pi$, and $\mathcal{N}(\bolds{\mu},\Sigma)$
stands for Gaussian distribution with mean $\bolds{\mu} \in
\mathbb{R}^p$ and covariance matrix $\Sigma$. For simplicity, we
consider $\pi=\frac{1}{2}$ and the 0--1 loss.
We will define\vadjust{\goodbreak} the ratio of separation as a measure of the fraction
of ``information'' carried by the subset of features due to thinning
with respect to that of the original feature set and show that this
quantity is ``preserved'' upon thinning. For simplicity, we take
$J=2$ (i.e., random splits of $\mathcal{F}$) and similar discussion
applies to $J
> 2$.

Let the feature set $\mathcal{F}$ be decomposed as
%
\begin{equation}
\label{eqrandomSplit} \mathcal{F}=\mathcal{F}_1 \cup\mathcal{F}_2\qquad
\mbox{such that } \mathcal{F}_1 \cap\mathcal{F}_2 =\varnothing
\mbox{ and } |\mathcal{F}_1|=\frac{p}{2} \triangleq m.
\end{equation}
We will show that each of the two subsets of features,
$\mathcal{F}_1$ and $\mathcal{F}_2$, carries a substantial fraction
of the ``information'' contained in the original data when $p$ is
large, assuming the data is generated from Gaussian mixture
\eqref{eqgm}.

A quantity that is crucial in our inquiry is
%
\begin{equation}
\label{defseparation} S_{\mathcal{F}}
=\mathbf{u}_{\mathcal{F}}^T\Sigma^{-1}_{\mathcal{F}}\mathbf{u}_{\mathcal{F}},
\end{equation}
where $\mathbf{u}_{\mathcal{F}}=(\bolds{\mu}_1-\bolds{\mu}_2)_{\mathcal{F}}
\triangleq(U_1,U_2,\ldots,U_p)_{\mathcal{F}}$ and here $\mathcal{F}$,
as a subscript, indicates that the associated quantity corresponds
to the feature set $\mathcal{F}$. We call~$S_{\mathcal{F}}$ the
separation of the Gaussian mixture induced by the feature set~$\mathcal{F}$.
The separation is closely related to the Bayes error
rate for classification through a well-known result in multivariate
statistics.
\begin{lemma}[{[\citet{Anderson1958}]}]
For Gaussian mixture \eqref{eqgm} and 0--1 loss, the Bayes error
rate is given by
$\Phi(-\frac{1}{2}(\mathbf{u}_{\mathcal{F}}^T\Sigma^{-1}_{\mathcal{F}}\mathbf
{u}_{\mathcal{F}})^{1/2})$
where $\Phi(\cdot)$ is defined as $\Phi(x)=\int_{-\infty}^{x}
\frac{1}{\sqrt{2\pi}}e^{-{z^2}/{2}}\,dz$.
\end{lemma}

Let the covariance matrix $\Sigma$ be written as
\[
\Sigma= \left[
\matrix{
A & B^T\vspace*{2pt}\cr
B & D}\right],
\]
where we assume block $A$ corresponds to features in
$\mathcal{F}_1$ after a permutation of rows and columns.
Accordingly, write $\mathbf{u}$ as
$\mathbf{u}_{\mathcal{F}}=(\mathbf{u}_{\mathcal{F}_1},
\mathbf{u}_{\mathcal{F}_2})$ and define $S_{\mathcal{F}_1}$ (called the
separation induced by $\mathcal{F}_1$) similarly as
\eqref{defseparation}. Now we can define the \textit{ratio of
separation} for the feature subset $\mathcal{F}_1$ as
%
\begin{equation}
\label{definfoFraction}
\gamma=\frac{S_{\mathcal{F}_1}}{S_{\mathcal{F}}}.
\end{equation}
%

To see why definition \eqref{definfoFraction} is useful, we give
here a numerical example. Assume there is a Gaussian mixture defined
by \eqref{eqgm} such that $\Sigma_{100 \times100}$ is
a~tri-diagonal matrix with diagonals being all $1$ and off-diagonals
being $0.6$, $\mathbf{u}_{\mathcal{F}}=(1,\ldots,1)^T$. Suppose one picks
the first $50$ variables and form a new Gaussian mixture with
covariance matrix $A$ and mixture center distance~$\mathbf{u}_{\mathcal{F}_1}$.
We wish to see how much is affected in
terms of the Bayes error rate. We have
\begin{eqnarray*}
S_{\mathcal{F}}&=&45.87,\qquad
\Phi\bigl(-\tfrac{1}{2}(\mathbf{u}_{\mathcal{F}}^T\Sigma^{-1}_{\mathcal{F}}\mathbf
{u}_{\mathcal{F}})^{1/2}\bigr)=3.54
\times
10^{-4},\\
 S_{\mathcal{F}_1}&=&23.32,\qquad
\Phi\bigl(-\tfrac{1}{2}({\mathbf{u}_{\mathcal{F}_1}}^T
A^{-1}{\mathbf{u}_{\mathcal{F}_1}})^{1/2}\bigr)=7.87 \times10^{-3}
\end{eqnarray*}
and $\gamma=0.5084$. Here the difference between feature set
$\mathcal{F}_1$ and $\mathcal{F}$ is very small in terms of their
classification power. In general, if the dimension is sufficiently
high and $\gamma$ is nonvanishing, then using a subset of features
will not incur much loss in classification power. In
Theorem~\ref{thmsepPreserve}, we will show that, under certain
conditions, $\gamma$ does not vanish (i.e., $\gamma>c$ for some
positive constant $c$) so a feature subset is as good as the whole
feature set in terms of classification power.

Our main assumption (i.e., in Theorem~\ref{thmsepPreserve}) is
actually a technical one related to the ``local'' dependency among
components of $\mathbf{u}$ after some variable transformation. The exact
context will become clear later in the proof of
Theorem~\ref{thmsepPreserve}. For now, let $\Sigma$ have a Cholesky
decomposition $\Sigma=HH^T$ for some lower triangular matrix~$H$. A
variable transformation in the form of
$\mathbf{y}=H^{-1}\mathbf{u}_{\mathcal{F}}$ will be introduced. The idea is
that we desire $\Sigma=HH^T$ to possess a structure such that the
components of $\mathbf{y}=H^{-1}\mathbf{u}_{\mathcal{F}}$ are ``locally''
dependent so that some form of law of large numbers may be applied.
To avoid technical details in the main text, we shall discuss the
assumption in the supplement [\citet{refsupp}].

Our main result is the following theorem. 
\begin{theorem}
\label{thmsepPreserve} Assume the data are generated from Gaussian
mixture~\eqref{eqgm}. Further assume the smallest eigenvalue of
$\Sigma^{-1}$, denoted by $\lambda_{\mathrm{min}}(\Sigma^{-1})$, is bounded
away from $0$ under permutations of rows and columns of $\Sigma$.
Then, under assumptions of Lemma~3 (c.f. Supplement A [\citet{refsupp}]),
the separation induced by the feature set $\mathcal{F}_1$ satisfies
\[
\frac{S_{\mathcal{F}_1}}{S_{\mathcal{F}}} \geq
\biggl(\frac{1}{2}\biggr)^{-}
\]
in probability as $p \rightarrow\infty$ where $(a)^{-}$ indicates
any constant smaller than $a$. When the number of partitions $J >
2$, the right-hand side is replaced by $1/J$.
\end{theorem}
\begin{pf}
See supplement [\citet{refsupp}] for proof.
\end{pf}


\subsubsection{Related work} \label{sectionthinningRelated} There
are mainly two lines of work closely related to ours. One is the
Johnson--Lindenstrauss lemma and related
[\citet{JohnsonLindenstrauss1984}, \citet{DasguptaGupta2002}]. The
Johnson--Linden\-strauss (or J--L) lemma states that, for Gaussian
mixtures in high-dimensional space, upon a random projection to a
low-dimensional subspace, the separation between the mixture centers
in the projected space is ``comparable'' to that in the original
space with high probability. The difference is that the random
projection in J--L is carried out via a nontrivial linear
transformation and the separation is defined in terms of the
Euclidean distance whereas, in our work, random projection is
performed coordinate-wise in the original space and we define the
separation with the Mahalanobis distance.

The other related work is the random subspace method
[\citet{TinHo1998}], an early variant of the RF classifier ensemble
algorithm, that is, comparable to bagging and Adaboost in terms of
empirical performance. The random subspace method grows a tree by
randomly selecting half of the features and then constructs a
tree-based classifier. However, beyond simulations there has been no
formal justification for the random selection of half of the
features. Our result provides support on this aspect. In a high
dimensional data setting where the features are ``redundant,'' our
result shows that a~randomly-selected half of the features can lead
to a tree comparable, in terms of classification power, to a
classifier that uses all the features; meanwhile the random nature
of the set of features used in each tree makes the correlation
between trees small, so good performance can be expected.

Our theoretical result, when used in co-training, can be viewed as a
manifestation of the ``blessings of the dimensionality''
[\citet{Donoho2000CurseBless}]. For high-dimensional data analysis,
the conventional wisdom is to do dimension reduction or projection
pursuit. As a result, the ``redundancy'' among the features is
typically not used and, in many cases, even becomes the nuisance one
strives to get rid of. This is clearly a waste. When the
``redundancy'' among features is complementary, such redundancy
actually allows one to construct two coupling learners from which
co-training can be carried out. It should be emphasized that the
splitting of the feature set works because of redundancy. We believe
the exploration of this type of redundancy will have important
impact in high-dimensional data analysis.

\section{Applications on TMA images}
\label{sectionexperiments} To assess the performance of TACOMA, we
evaluate a collection of TMA images from the Stanford Tissue
Microarray Database, or STMAD [see \citet{Marinelli07} and
\href{http://tma.stanford.edu/}{http://tma.stanford.}
\href{http://tma.stanford.edu/}{edu/}]. TMAs corresponding to the
potential expression of the estrogen receptor (ER) protein in breast
cancer tissue are used since ER is a histologically well-studied
marker that is expressed in the cell nucleus. An example of TMA
images can be seen in Figure~\ref{figureglcms}. There are $641$ TMA
images in this set and each image has been assigned a score from
$\{0,1,2,3\}$. 
The scoring criteria are as follows: ``$0$'' representing a definite negative
(no staining of cancer cells), ``$3$'' a definitive positive (a~majority of cancer cells show dark nucleus staining) and ``$2$'' for
positive (a~minority of cancer cells show nucleus staining or a
majority show weak nucleus staining). The score ``$1$'' indicates
ambiguous weak staining in a minority of cancer cells. The class
distribution of the scores is $(65.90\%,2.90\%,7.00\%,24.20\%)$.
Such an unbalanced class distribution makes the scoring task even
more challenging. In our experiments, we assess the accuracy by the
proportion of images in the test sample that receives the same score
by a scoring algorithm as that given by the reference set (e.g.,
STMAD). We split the images into a training and a test set of sizes
$313$ and $328$, respectively. In the following, we will first
describe\vadjust{\goodbreak} our choice of various design options and parameters, and
then report performance of TACOMA on a large training set (i.e., a set
of $313$ TMA images) and small training set (i.e., a set of $30$ TMA
images) with co-training in Section~\ref{subsectionexpScore} and
Section~\ref{subsectionexpCotraining}, respectively.

We use three spatial relationships, $(\nearrow,3)$, $(\searrow,1)$
and $(\nearrow,1)$, in our experiments. In particular, $(\nearrow,3)$
is used in our experiment on TACOMA for a large training set, while
$(\searrow,1)$ is used along with $(\nearrow,3)$ in our co-training
experiment and $(\nearrow,1)$ in an additional experiment (see
Table~\ref{tableaccuracy3Other}). Often $(\nearrow,1)$ is the
default choice in applications; we use $(\nearrow,3)$ here to
reflect the granularity of the visual pattern seen in the images.
Indeed, as the staining patterns in TMA images for ER markers occur
only in the nucleus, $(\nearrow,3)$ leads to a slightly better scoring
performance than $(\searrow,1)$ according to our experiment on
TACOMA; moreover, no significant difference is observed for TACOMA
when simply concatenating features derived from $(\nearrow,3)$ and
$(\searrow,1)$. $(\searrow,1)$ is used in co-training in hoping that
it is less correlated to features derived from $(\nearrow,3)$ than
others, as these two are on the orthogonal directions.

For a good balance of computational efficiency, discriminative
power, as well as ease of implementation, we take $N_g=51$ (our
experiments are not particularly sensitive to the choice of $N_g$ in
the range of $40$ to $60$) and apply uniform quantization over the
$256$ gray levels in our application. One can, of course, further
explore this, but we would not expect a substantial gain in
performance due to the limitation of a computer algorithm (or even
human eyes) in distinguishing subtle differences in gray levels
given a moderate sample size.

We use the R package ``randomForest''\footnote{Originally written
in Fortran by Leo Breiman and Adele Cutler, and later ported to~R by
Andy Liaw and Matthew Wiener.} in this work. There are two important
parameters, the number of trees in the ensemble and the number of
features to explore at each node split. These are searched through
$\{50, 100, 200, 500\}$ and $\{0.5\sqrt{p},\sqrt{p}, 2\sqrt{p}\}$
($\sqrt{p}$ is the default value suggested by the R package for $p$
the number of features fed to RF), respectively, in this work and
the best test set error rates are reported. More information on RF
can be found in \citet{RF}.

\subsection{Performance on large training set}
\label{subsectionexpScore} The full set of $313$ TMA images in the
training set are used in this case. We run TACOMA on the training
set (scores given by STMAD) and apply the trained classifier to the
test set images to obtain TACOMA scores. Then, we blind STMAD scores
in the test set of $328$ images ($100$ of which are duplicated so
totally there are $428$ images) and have them reevaluated by two
experienced pathologists from two different institutions. The $100$
duplicates allow us to evaluate the self-consistency of the
pathologists.\vadjust{\goodbreak}

Although the scores from STMAD do not necessarily represent ground
truth, they serve as a fixed standard with respect to which the
topics of accuracy and reproducibility can be examined. On the test
set of $328$ TMA images, TACOMA achieves a classification accuracy
of $78.57\%$ (accuracy defined as the proportion of images receiving
the same score as STMAD). We argue this is close to the optimal. The
Bayes rate is estimated for this particular data example
(represented as GLCMs) with a simulation using a $1$-nearest
neighbor ($1\mathit{NN}$) classifier. The Bayes rate refers to the
theoretically best classification rate given the data distribution.
With the same training and test sets as RF classification, the
accuracy achieved by $1\mathit{NN}$ is around $60\%$. According to a
celebrated theorem of \citet{CoverHart1967}, the error rate by $1\mathit{NN}$
is at most twice that of the Bayes rule. This result implies an
estimate of the Bayes rate at around $80\%$ subject to small sample
variation (the estimated Bayes rates on the original image or its
quantized version are all bounded above by this number according to
our simulation). Thus, TACOMA is close to optimal.

\begin{figure}[b]
\vspace*{-2pt}
\includegraphics{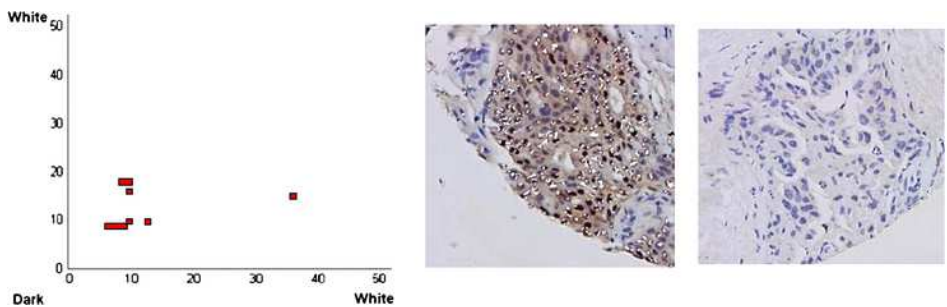}

\caption{The salient pixels (highlighted
in white). The left panel displays top features (indices of
GLCM entries) from the classifier where the $x$-axis and $y$-axis
indicate the row and column of the GLCM entries. The middle and
right panels display images having scores 3 and 0, respectively; the
pixels highlighted in white are those that correspond to the GLCM
entries shown in the left panel. Note the highlighted pixels in the
right panel are notably absent. For visualization, only part of the
images are shown (see Supplement B [\protect\citet{refsupp}] for larger images).}
\label{figuresalientfeaturepixels}
\end{figure}

In the above experiments, we use four image patches. To see if
TACOMA is sensitive to the choice of image patches, we conduct
experiments over a~range of different patch sets and achieve an
average accuracy at $78.66\pm0.52\%$. Such an accuracy indicates
that TACOMA is robust to the choices of image patches.

It is worth reemphasizing that all reports of test errors for TMA
images are not based on absolute truth, as all scores given to these
images are subjectively provided by a variable human scoring
process.

\subsubsection*{Salient spots detection} The ability of TACOMA to
detect salient pixels is demonstrated in
Figure~\ref{figuresalientfeaturepixels}\vadjust{\goodbreak} where image pixels are
highlighted in white if they are associated with a significant
scoring feature. These highlighted pixels are verified by the
pathologists to be indicative. With relatively few exceptions, these
locations correspond to areas of stained nuclei in cancer cells. We
emphasize that these highlighted pixels indicate features most
important for classification as opposed to identifying every
property indicative of ER status. The highlighted pixels facilitate
interpretation and the comparison of images by
pathologists.

\begin{figure}

\includegraphics{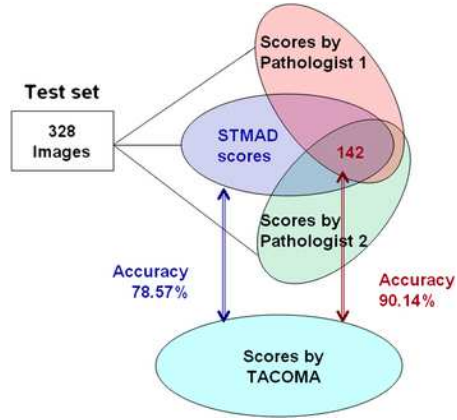}

\caption{Classification performance of
TACOMA. On the STMAD test set TACOMA achieves an accuracy of
$78.57\%$. On the $142$ images assigned a unanimous score by two
pathologists and STMAD, TACOMA agrees on about $90\%$.}
\label{figurevennLKS}
\end{figure}

\subsubsection*{The experiments with pathologists}
The superior classification performance of TACOMA is also
demonstrated by scores provided by the two pathologists. These two
copies of scores, along with STMAD, provide three independent
pathologist-based scores. Among these, $142$ images receive\break a~unanimous score.
Consequently, these may be viewed as a reference
set of ``true'' scores against which the accuracy of TACOMA might be
evaluated (accuracy being defined as the proportion of images
receiving the same score as the reference set). Here, TACOMA
achieves an accuracy of $90.14\%$; see Figure~\ref{figurevennLKS}.\vadjust{\goodbreak}

Scores provided by the two pathologists are also used to assess
their self-consistency. Here self-consistency is defined as the
proportion of repeated images receiving an identical score by the
same pathologist. Consensus among different pathologists is an issue
of concern [\citet{PennaGrill1997}, \citet{Walker06}]. In order to obtain
information about the self-consistency of pathologist-based scores,
$100$ images are selected from the set of $328$ images. These $100$
images are rotated and/or inverted, and then mixed at random with
the $328$ images to avoid recognition (so each pathologist actually
scored a total of $428$ TMA images). The self-consistency rates of
the two pathologists are found to\vadjust{\goodbreak} be in the range 75--84\%. Of
course, one desirable feature of any automated algorithm such as
TACOMA is its complete (i.e., $100\%$) self-consistency.

\subsubsection*{Performance comparison of RF to SVM and boosting}
The classification algorithm chosen for TACOMA is RF. Some popular
alternatives include support vector machines (SVM)
[\citet{CortesVapnik1995}], Boosting [\citet{FreundSchapire1996}] and
Bayesian network [\citet{Pearl1985}], etc. Using the same training and
test set as that for RF, we conduct experiments with SVM as well as
Boosting of a naive Bayes classifier. The input features for both
SVM and Boosting are the entries of the GLCM. We use the Libsvm
[\citet{LIBSVM}] software for SVM. The naive Bayes classifier is
adopted from \citet{ASPRAS2006}. The idea is to find the class that
maximizes the posterior probability for a new observation, denoted
by $(I(x)=a_x, x\in\{1,2,\ldots,N_g\}\otimes\{1,2,\ldots,N_g\})$ for
a fixed $N_g$. That is, we seek to solve
\[
\arg\max_{k \in\{0,1,2,3\}}
\operatorname{Prob} \{k|I(1,1)=a_{1,1},\ldots,I(N_g,N_g)=a_{N_g,N_g}\}
\]
under the assumption that $(I(1,1),\ldots,I(N_g,N_g))$ follows a
multinomial distribution. More details can be found in
\citet{ASPRAS2006}. The results are shown in
Table~\ref{tablecomparisonRFSVMBoost}. We can see that RF
outperforms SVM and Boosting by a large margin. This is consistent
with observations made by \citet{HolmesKapelner2009}.

\subsection{Experiments on small training sets}
\label{subsectionexpCotraining} We conduct experiments on
co-training with natural splits and thinning. For natural splits, we
use GLCM's corresponding to two spatial relationships,
$(\nearrow,3)$ and $(\searrow,1)$, as features. For thinning, we
combine features corresponding to $(\nearrow,3)$ and $(\searrow,1)$
and then split this combined feature set.

The number of labeled examples is fixed at $30$ (compared to $313$
in experiments with a large training set). This choice is designed to
make it easy to get a nonempty class $1$ (which carries only about
$2.90\%$ of the cases). We suspect this number can be further
reduced without suffering much in learning accuracy. The test set is
the same as that in Section~\ref{subsectionexpScore}. The result is
shown in Table~\ref{tableexpCotrainingTMA}.
One interesting observation is that co-training by thinning
achieves an accuracy very close to that by natural splits.
Additionally, Table~\ref{tableexpCotrainingTMA} lists error rates
given by RF on features corresponding to $(\nearrow,3) \cup
(\searrow,1)$ and its thinned subsets. Here thinning of the feature
set does not cause much loss in RF performance, consistent with our
discussion in Section~\ref{sectioncotrainingTheory}.

%
\begin{table}
\caption{Performance of RF and co-training by thinning on TMA
images. The unlabeled set~is~taken as the test set in
Section~\protect\ref{subsectionexpScore} and the labeled~set is randomly
sampled from~the~corresponding training set. The subscript for
``thinning'' indicates the~number~of~partitions. The results are
averaged over $100$ runs and over~the~two~coupling classifiers for
co-training}\label{tableexpCotrainingTMA}
\begin{tabular*}{\textwidth}{@{\extracolsep{\fill}}lc@{}}
\hline
\textbf{Scheme} &\textbf{Error rate} \\
\hline RF on $(\nearrow,3) \cup(\searrow,1)$ &34.36\% \\
$\mbox{Thinning}_2$ on $(\nearrow,3) \cup(\searrow,1)$ &34.21\% \\
$\mbox{Thinning}_3$ on $(\nearrow,3) \cup(\searrow,1)$ &34.18\% \\
[3pt]
Co-training by natural split on $(\nearrow,3)$ and $(\searrow,1)$
&27.49\% \\
Co-training by $\mbox{thinning}_2$ on $(\nearrow,3) \cup(\searrow,1)$
&27.89\% \\
Co-training by $\mbox{thinning}_3$ on $(\nearrow,3) \cup(\searrow,1)$
&27.62\% \\
\hline
\end{tabular*}
\end{table}

\subsection{Experiments of TACOMA on additional data sets}\label{sec4.3} This
study focuses on the ER marker for which the staining is nuclear.
However, the TACOMA algorithm can be applied with equal ease to
markers that exhibit cell surface,\vadjust{\goodbreak} cytoplasm or other staining
patterns. Additional experiments are conducted on the Stanford TMA
images corresponding to three additional protein markers: CD117,
CD34 and NMB. These three sets of TMA images are selected for their
large sample size and relatively few missing scores (excluded from
experiment). The results are shown in
Table~\ref{tableaccuracy3Other}. In contrast, the automated scoring
of cytoplasmic markers is often viewed as more difficult and refined
commercial algorithms for these were reportedly not available in a
recent evaluation [\citet{CampNeuRimm08}] of commercial scoring
methods.

%
\begin{table}[b]
\caption{Accuracy of TACOMA on TMA images corresponding to
protein markers CD117, CD34,~NMB and ER. Except for ER (which has a
fixed training and test set), we~use~$(\nearrow,1)$~and~80\% of the
instances for training and the rest for test; this~is~repeated~for
$100$ runs and results averaged}\label{tableaccuracy3Other}
\begin{tabular*}{\textwidth}{@{\extracolsep{\fill}}lccc@{}}
\hline
\textbf{Marker} & \textbf{Staining} &\textbf{\#Instances} &\textbf{Accuracy}\\
\hline
ER &Nucleus &\phantom{0}641 &$78.57\%$\\
CD117 &Cell surface &1063 &81.08\%\\
NMB &Cytoplasmic &1036 &84.17\%\\
CD34 &Cytoplasmic and cell surface &\phantom{0}908 &76.44\%\\
\hline
\end{tabular*}
\end{table}


\section{Discussion}
\label{sectionconclusion} We have presented a new algorithm that
automatically scores TMA images in an objective, efficient and
reproducible manner. Our contributions include the following: (1) the
use of
co-occurrence counting statistics to capture the spatial regularity
inherent in a heterogeneous and irregular set of TMA images; (2) the
ability to report salient pixels in an image that determine its
score; (3) the incorporation of pathologists' input via informative
training patches which makes our algorithm adaptable to various
markers and cell types; (4) a very small training sample is
achievable with co-training and we have provided some theoretical
insights into co-training via thinning of the feature set. Our
experiments show that TACOMA can achieve performance comparable to
well-trained pathologists. It uses the similar set of pixels for
scoring as that would be used by a pathologist and is not adversely
sensitive to the choice of image patches. The theory we have
developed on the thinning scheme in co-training gives insights on
why thinning may rival the performance of a natural split in
co-training; a thinned slice may be as good as the whole feature set
in terms of classification power, hence, thinning can lead to two
strong coupling classifiers that will be used in co-training and
this is what a natural split may achieve.

The utility of TACOMA lies in large population-based studies that
seek to evaluate potential markers using IHC in large cohorts. Such
a study may be compromised by a scoring process, that is, protracted,
prohibitively expensive or poorly reproducible. Indeed, a manual
scoring for such a study could require hundreds of hours of
pathologists' time without achieving a~reproducibly consistent set
of scores. Experiments with several IHC markers demonstrate that
our approach has the potential to be as accurate as manual scoring
while providing a fast, objective, inexpensive and highly
reproducible alternative. Even more generally, TACOMA may be adopted
to other types of textured images such as those appearing in remote
sensing applications. These properties provide obvious advantages
for any subsequent statistical analysis in determining the validity
or clinical utility of a potential marker. Regarding
reproducibility, we note that the scores provided by two
pathologists in our informal pilot study revealed an intra-observer
agreement of around 80\% and an accuracy only in the range of 70\%,
as defined by the STMAD reference set (excluding all images deemed
unscorable by the pathologists). This low inter-observer agreement
may be attributed to a variety of factors, including a lack of a
subjective criteria used for scoring or the lack of training against
an established standard. This performance could surely be improved
upon, but it highlights the inherent subjectivity and variability of
human-based scoring.

In summary, TACOMA provides a transparent scoring process that can
be evaluated with clarity and confidence. It is also flexible with
respect to marker patterns of cellular localization: although the ER
marker is characterized by staining of the cell nucleus, TACOMA
applies with comparable ease and success to cytoplasmic or other
marker staining patterns (see Table~\ref{tableaccuracy3Other} in Section~\ref{sec4.3}).

A software implementation of TACOMA is available upon request and
the associated R package will be made available to the R project.

\section*{Acknowledgments} The authors would like to thank the
Associate Editor and the anonymous reviewers for their constructive
comments and suggestions.

%
\begin{supplement}[id=suppA]
\stitle{Supplement A: Assumption $\mathcal{A}_1$, proof of Theorem~\ref{thmsepPreserve}
and simulations on thinning}
\slink[doi]{10.1214/12-AOAS543SUPPA}
\slink[url]{http://lib.stat.cmu.edu/aoas/543/supplementA.pdf}
\sdatatype{.pdf}
\sdescription{We provide a detailed description of Assumption
$\mathcal{A}_1$, a sketch of the proof of Theorem~\ref{thmsepPreserve} and simulations
on the ratio of separation upon thinning under different settings.}
\end{supplement}
%


\begin{supplement} [id=suppB]
\stitle{Supplement B: TMA images with salient pixels marked\\}
\slink[doi]{10.1214/12-AOAS543SUPPB}
\slink[url]{http://lib.stat.cmu.edu/aoas/543/supplementB.pdf}
\sdatatype{.pdf}
\sdescription{This supplement contains a close view of some TMA images
where the
salient pixels are highlighted.}
\end{supplement}


%


\printaddresses

\end{document}